\documentclass[11pt]{article}

\usepackage[a4paper,margin=1in]{geometry}
\usepackage[T1]{fontenc}
\usepackage[utf8]{inputenc}
\usepackage{lmodern}
\usepackage{microtype}
\usepackage{setspace}
\usepackage{graphicx}
\graphicspath{{figures/}}
\usepackage{float}
\usepackage[font=small,labelfont=bf,labelsep=period,justification=centering]{caption}

\usepackage{booktabs}
\usepackage{array}
\usepackage{tabularx}
\usepackage{longtable}
\newcolumntype{L}[1]{>{\raggedright\arraybackslash}p{#1}}

\usepackage{enumitem}
\setlist{itemsep=2pt,topsep=4pt}
\usepackage{amsmath,amssymb}

\usepackage{listings}
\lstdefinelanguage{json}{
  basicstyle=\ttfamily\footnotesize,
  showstringspaces=false,
  breaklines=true,
  frame=single,
  framerule=0.3pt,
  rulecolor=\color{black!30},
  xleftmargin=0.5em,
  xrightmargin=0.5em,
  literate=
    *{:}{{{\color{black}{:}}}}{1}
     {,}{{{\color{black}{,}}}}{1}
     {\{}{{{\color{black}{\{}}}}{1}
     {\}}{{{\color{black}{\}}}}}{1}
     {[}{{{\color{black}{[}}}}{1}
     {]}{{{\color{black}{]}}}}{1}
}
\usepackage{xcolor}

\usepackage[explicit]{titlesec}
\titleformat{\section}{\Large\bfseries}{\thesection}{0.6em}{#1}
\titleformat{\subsection}{\large\bfseries}{\thesubsection}{0.5em}{#1}
\titleformat{\subsubsection}{\normalsize\bfseries}{\thesubsubsection}{0.4em}{#1}
\titlespacing*{\section}{0pt}{1.4\baselineskip}{0.6\baselineskip}
\titlespacing*{\subsection}{0pt}{1.0\baselineskip}{0.4\baselineskip}
\titlespacing*{\subsubsection}{0pt}{0.8\baselineskip}{0.3\baselineskip}

\usepackage[hidelinks,breaklinks=true,colorlinks=false]{hyperref}
\usepackage{url}
\renewenvironment{abstract}
  {\begin{center}\bfseries\large Abstract\end{center}%
   \begin{quotation}\small\noindent\ignorespaces}
  {\end{quotation}}

\title{\bfseries Kettle: Attested Builds for Verifiable Software Provenance}
\author{%
  \begin{tabular}{c@{\hskip 2em}c}
    Andr\'e Arko & Amean Asad \\
    \texttt{andre@confidential.ai} & \texttt{amean@confidential.ai}
  \end{tabular}
  \\[1.4em]
  \href{https://confidential.ai}{Confidential.ai}
}
\date{April 2026}

\begin{document}
\maketitle

\begin{abstract}
Kettle is an \textbf{attested build} system that produces cryptographically verifiable provenance for software built inside Trusted Execution Environments (TEEs). A Kettle build records the source commit, dependency set, toolchain, build environment, and output artifact digests in a provenance document produced inside a measured confidential VM. The SHA-256 digest of that document is committed to the TEE platform's attestation report-data field, so the hardware-signed attestation report is itself the signature on the provenance, with the signing identity chaining to the TEE manufacturer's root of trust rather than to the build infrastructure operator. Because the CVM image is itself reproducible, its launch measurement is public and stable, which lets a build requester pre-attest the CVM before submitting any input and optionally deliver source over a TLS channel terminated inside it, so the build runs end-to-end confidentially without the host ever seeing source code in plaintext. Verification reduces to one signature check against the vendor root and a small set of digest comparisons, with no need to re-execute the build. The result removes the build infrastructure, its operators, and the artifact distribution channel from the trust surface a verifier must accept when deciding whether a binary corresponds to its claimed inputs.
\end{abstract}

\section{Introduction}

\subsection{Motivation}

Software consumers routinely execute artifacts built by a separate software provider. Examples include a developer installing a prebuilt binary from a registry, an enterprise deploying a vendor container image, or a customer running a hosted service whose source they were allowed to inspect.

A software artifact is the end result of a rather complicated build system. Developers write source code, and that source is fetched from version control. Dependencies are resolved from package registries or caches. Compilers, linkers, and build scripts transform those inputs while running with specific environment variables and filesystem state. The operating system, kernel, and container or VM image define the execution environment in which the build occurs. Changes to any of these layers affect the bytes that are eventually distributed, and hence impact the behavior of the software that a consumer interacts with. Each of these layers also presents an attack surface.

The software provider then must provide some claims regarding the integrity and behavior of the artifacts it produces. A consumer then uses those claims to establish trust in the software they interact with from that provider. The usual evidence available to that consumer is operational: CI logs, release notes, signed checksums, artifact registry metadata, or a provenance document emitted by the build pipeline. These records can be useful, but by themselves they prove only that some system made a statement. They do not prove that the statement is true. A compromised build runner can inject code during compilation while leaving the source repository clean. A privileged operator can alter the toolchain, cached dependencies, environment variables, or artifact upload step. An attacker with access to artifact storage can replace the binary after the build has finished. In each case the consumer sees a plausible release record, but the artifact may no longer correspond to the source and dependencies being claimed.

The underlying gap is that most build systems do not bind the provider's claims to the actual software artifact in a way that is independently provable. A release may point to a source commit, publish a lockfile, name a toolchain, and distribute an artifact digest, but those facts are usually connected by process rather than by cryptographic evidence. Closing this gap is difficult because build systems are large, stateful, and highly variable across languages, package managers, operating systems, and deployment formats. Any useful binding must scale across that complexity, and verification must remain fast and low-friction enough for ordinary software consumption.

\subsection{Problem Statement}
\label{sec:problem-statement}

The problem is to establish, for a specific software artifact, whether the provider's claim about its origin is trustworthy. The aim is to turn build provenance from an assertion into cryptographic evidence that a consumer can verify independently.

Any practical solution has to satisfy the following constraints:

\begin{enumerate}
\item A software consumer, or tooling acting on the consumer's behalf, can evaluate the evidence without inspecting the provider's internal systems.
\item Within the threat model, evidence cannot be forged or silently altered by registries, mirrors, caches, or deployment systems without detection by verification.
\item The approach scales across large volumes of artifacts and consumers.
\item Verification is efficient enough for ordinary software consumption, not only for manual audits.
\item The approach tolerates different languages, dependency managers, toolchains, operating systems, and artifact formats.
\end{enumerate}

The desired outcome is build provenance that is cryptographically verifiable. For each artifact, the provider produces a record of how it was built (the source it came from, the dependencies and toolchain it used, the environment it ran in) and binds that record to the artifact with hardware-rooted cryptographic evidence. A consumer checks the evidence directly. If any of the relevant inputs, the build process, the environment, or the artifact itself differs from what the record states, verification fails.

\subsection{Scope}

This paper describes the threat model and architecture of Kettle, a build tool developed by Confidential.ai to create and verify attested builds. It covers the technologies that together make attested builds trustworthy, describes the evidence chain from source inputs to output artifacts, and presents Kettle as an open-source implementation of that system. It does not attempt to prove that source code is safe, that dependencies are free of backdoors, or that a compiler is semantically correct. Attestation proves provenance and build-process integrity, not software intent.

\section{Background}

\subsection{Software Builds}

In the context of this paper, a software build is any transformation of inputs into outputs. Building can include preprocessing, compilation, linking, copying, packaging, or any kind of file generation. The specific aspects of a build that matter when creating builds that offer high confidence that the stated inputs map to a given output are:

\begin{itemize}
\item Inputs: source repositories, dependencies, toolchains, configuration files
\item Process: build scripts run, tool options, environment variables
\item Environment: operating system, kernel and its parameters, filesystems and other devices
\item Outputs: binaries, containers, tarballs, or other resulting artifacts
\end{itemize}

\begin{figure}[H]
\centering
\includegraphics[width=0.85\linewidth]{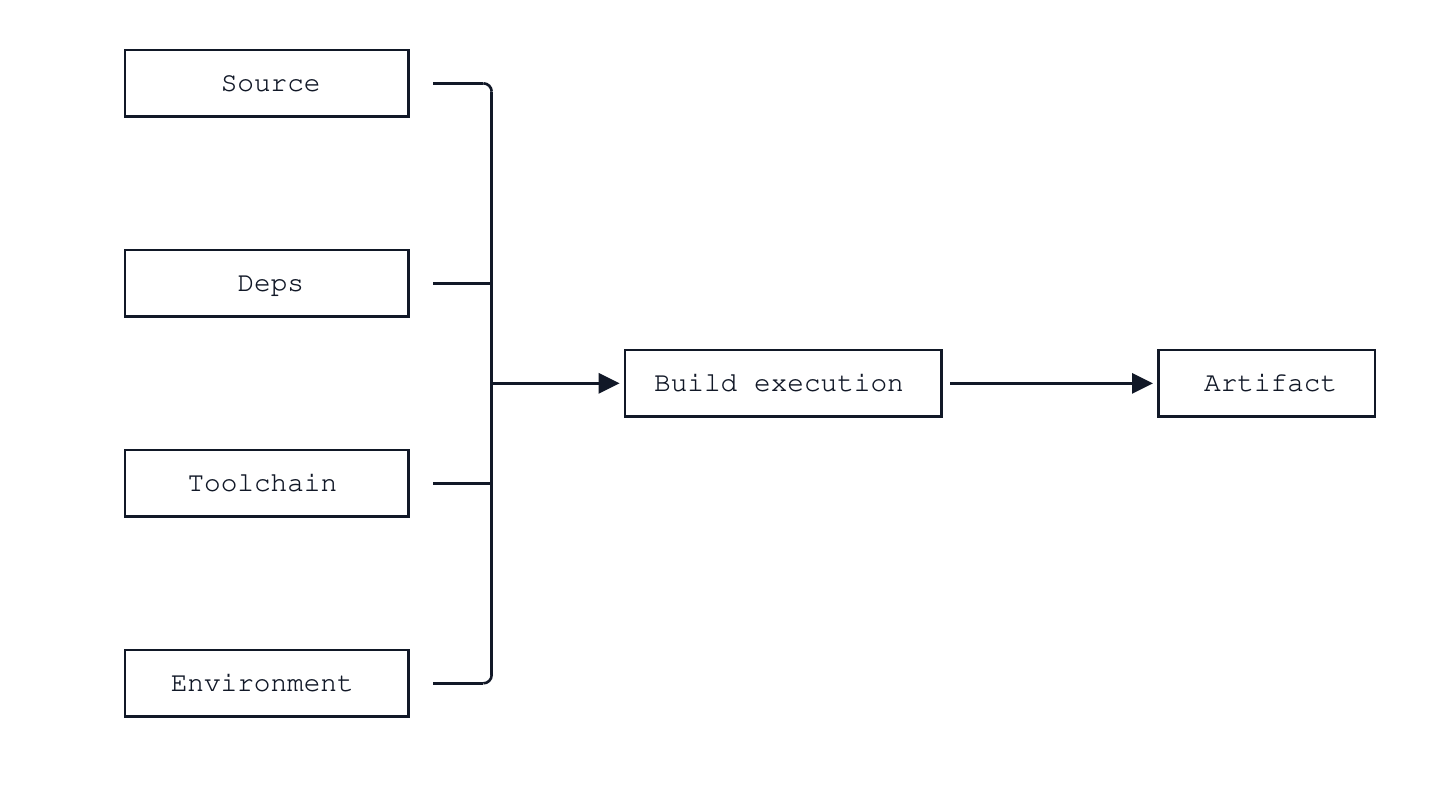}
\caption{A build is a many-input transformation. Source identity is necessary but incomplete because dependencies, toolchains, build scripts, and execution state also affect the final bytes.}
\label{fig:build-inputs}
\end{figure}

\subsection{SLSA Software Provenance}

As described by The Linux Foundation, ``SLSA is a specification for describing and incrementally improving supply chain security, established by industry consensus''~\cite{ref:slsa-provenance}. Among other things, the SLSA specification offers a concrete format for \textbf{provenance}, which is defined as verifiable information that can be used to track an artifact back, through all the moving parts in a complex supply chain, to where it came from. It is the verifiable information about software artifacts describing \textbf{where, when, and how} something was produced~\cite{ref:slsa-provenance}.

By using the SLSA v1.2 build provenance format, Kettle is able to provide industry-standard and machine-readable information that connects an artifact back to the source and tools that were used to create that artifact.

\subsection{Trusted Execution Environments (TEEs)}

A Trusted Execution Environment (TEE) is a hardware-enforced isolation boundary that provides four security properties:

\begin{itemize}
\item \textbf{Confidentiality.} Private memory within the TEE is encrypted at the hardware level. Memory pages are encrypted with keys managed by the CPU's secure processor, so the hypervisor, host operating system, and DMA-capable devices cannot decrypt private TEE memory contents through software access.
\item \textbf{Integrity.} Once software is initialized in a TEE, modifications to protected guest state by external software are detected. Replay attacks, data corruption, and memory remapping trigger hardware-level faults within the platform's threat model.
\item \textbf{Verifiability.} The TEE produces a cryptographic measurement (a digest) of the software loaded at launch. This measurement covers firmware, kernel, initial filesystem state, and user space, and is computed by the CPU's secure processor before the guest operating system begins execution.
\item \textbf{Attestation.} The TEE generates signed attestation reports binding the launch measurement to guest-supplied report data. These reports are signed by keys that chain to the hardware manufacturer's root of trust and cannot be forged by any software on the host.
\end{itemize}

\begin{figure}[H]
\centering
\includegraphics[width=0.95\linewidth]{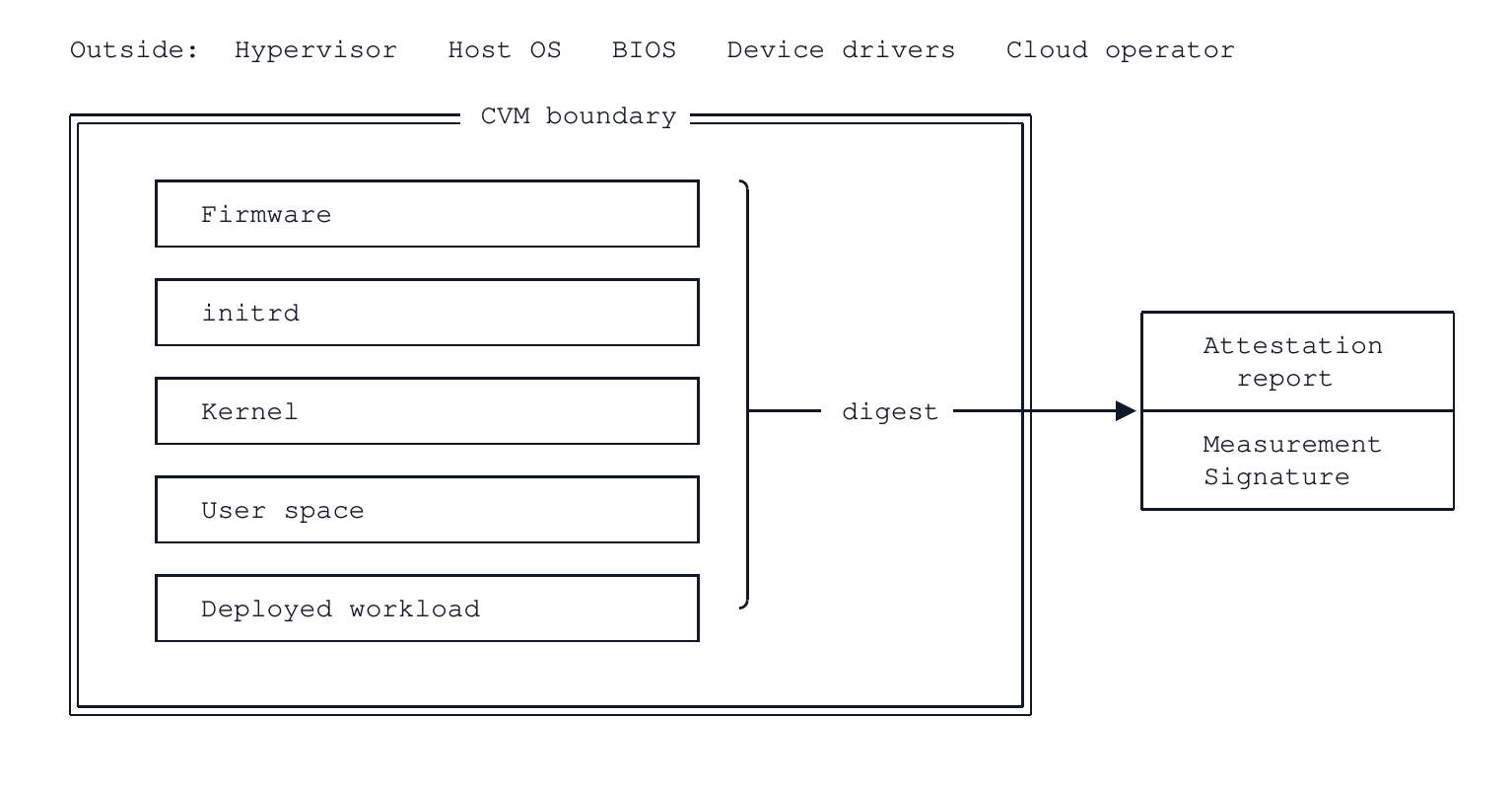}
\caption{Measured boot inside a CVM. The hypervisor, host OS, BIOS, device drivers, and cloud operator sit outside the confidentiality boundary. Inside, the security processor extends a digest of each loaded layer (firmware, initrd, kernel, user space, deployed workload) into a single cumulative measurement that is later carried in the attestation report.}
\label{fig:tee-measured-boot}
\end{figure}

\textbf{Implementations.} Current commercial TEEs from AMD~\cite{ref:amd-snp} and Intel~\cite{ref:intel-tdx} implement all four properties at the boundary of a virtual machine. The resulting environments are referred to as Confidential Virtual Machines (CVMs), where private guest memory is encrypted with keys the hypervisor never possesses. The four properties reinforce one another. Host software can no longer directly read private CVM memory or silently rewrite measured guest state, which prevents software-level data exfiltration and tampering by the host. As long as the hardware manufacturers can be trusted and firmware vulnerabilities can be patched as they are found, the trust surface contracts to the CPU vendor and the cryptographic primitives the platform relies on, replacing administrators, hypervisors, and host operating systems on the trusted side of the boundary.

\section{Prior Art}

\subsection{Reproducible Builds}

Reproducible builds address software provenance by making the build itself deterministic. A build is reproducible when the same source code, dependencies, toolchain, build instructions, and environment produce bit-for-bit identical output across separate machines. The goal is to make the relationship between declared inputs and output bytes behave like a stable one-to-one mapping for verification. A given input set should always produce the same artifact, and a matching artifact digest should identify the input set that produced it. Under that condition, a consumer does not need to trust the original build machine or its operator. The consumer, or some independent rebuilder, can rebuild the artifact and compare digests. If the locally produced digest matches the distributed artifact digest, the artifact corresponds to the published inputs.

\begin{figure}[H]
\centering
\includegraphics[width=0.85\linewidth]{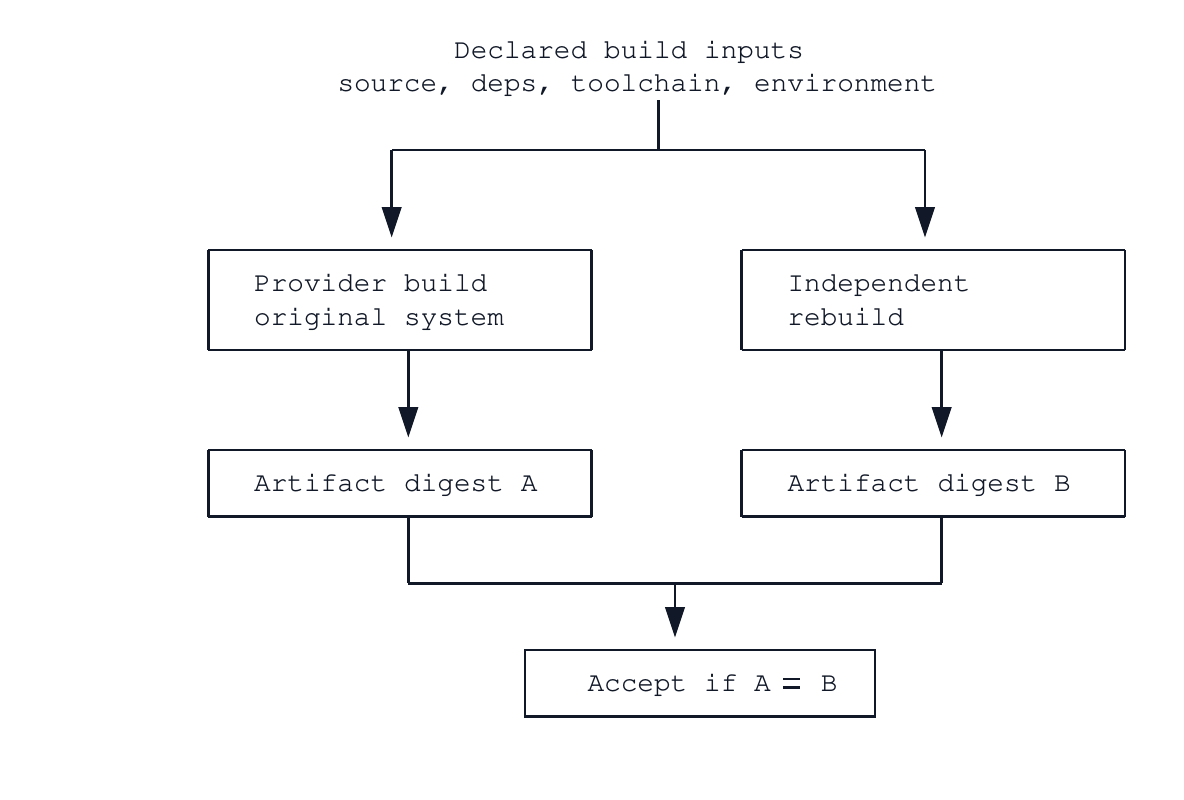}
\caption{Reproducible builds verify provenance by independent reconstruction. The original build and the rebuild must converge on the same output digest from the same declared inputs.}
\label{fig:reproducible-builds}
\end{figure}

A number of projects work on making this practical across the open source ecosystem. One of them is the Reproducible Builds project~\cite{ref:reproducible-builds}, which supports reproducibility efforts across many parts of that ecosystem. Lamb and Zacchiroli~\cite{ref:lamb-zacchiroli} survey the state of reproducibility across the Debian distribution and discuss its role in supply-chain integrity, including resistance to build-time tampering, independent verification of binaries, and reduced trust in build infrastructure.

The limitation is practical rather than conceptual. Many production builds are not deterministic. Compilers embed timestamps, parallel builds produce outputs in varying order, linkers record full file paths, archive tools preserve inconsistent ordering, and generated files may depend on host state. Achieving bit-for-bit identical output requires controlling these factors across the entire dependency tree. A single non-reproducible component anywhere in the chain breaks the guarantee.

Mapped against the constraints in \S\ref{sec:problem-statement}, this is where reproducible builds run into trouble. Achieving bit-for-bit determinism across an arbitrary set of languages, package managers, and toolchains is hard, which strains constraint 5 (toolchain coverage).

\subsection{Nix and Reproducible Build Environments}

Nix~\cite{ref:nix} is a package manager and build system that describes software builds as pure functions of their inputs, captured in files called derivations. Rather than requiring every build output to be bit-for-bit reproducible, Nix makes the build environment itself reproducible. A derivation or flake lock identifies the source, dependencies, toolchain, and build recipe used to construct an output, and Nix uses those identifiers to build the same software in the same way on any machine. This is a substantial improvement over package managers where dependency resolution can vary across machines or over time.

The distinction matters. A reproducible build environment does not by itself imply reproducible build output. The same pinned source and compiler can still produce different artifacts if the compiler, linker, archive tooling, generated files, or build scripts contain non-determinism. Nix can support reproducible builds when the underlying toolchain is deterministic, but the reproducibility property comes from the combination of a controlled environment and deterministic build behavior, not from environment pinning alone.

Nix also leaves a trust question at the build boundary. A derivation hash proves consistency with a particular Nix evaluation, but it does not prove that the evaluator, builder, or local toolchain executed honestly. A compromised build host can still alter the build process or output after evaluation. Nix therefore provides strong input and environment control, and can be part of a provenance strategy, but it does not by itself provide hardware-rooted evidence that a particular artifact was produced by a particular build execution.

\subsection{Attestable Builds}

An attestable build is one that runs inside a Trusted Execution Environment and emits hardware-signed evidence about the environment, inputs, and outputs observed during the build. Verification then becomes a check of that evidence (signature, environment measurement, and digest comparisons) rather than a re-execution of the build itself. The trust anchor moves from the build infrastructure and its operators to the TEE hardware and its attestation chain.

Two prior systems instantiate this idea concretely. \textbf{TEE Compile}, developed by Automata Network~\cite{ref:tee-compile}, runs a project's build inside an AWS Nitro Enclave. A worker process inside the enclave fetches dependencies through the host, builds the project, and emits a tuple of artifact bytes, an input/output hash report, and a Nitro attestation document whose PCR0 covers the enclave image. Verification compares input and output hashes against the report and checks that the enclave image is one the verifier trusts. The trust root is the AWS Nitro Attestation PKI rather than a CPU vendor's root, and Nitro Enclaves are constrained micro-VMs rather than general-purpose confidential VMs. How a verifier independently obtains a trustworthy reference measurement for the enclave image itself is not specified.

\textbf{Hugenroth, Lins, Mayrhofer, and Beresford}~\cite{ref:hugenroth} generalize the approach to AMD SEV-SNP, Intel TDX, and AWS Nitro Enclaves alike. Their design treats the host and the build process as untrusted, and introduces an integrity-protected Enclave Client inside the TEE. That client records the repository snapshot hash before the untrusted build starts, runs the build inside an inner sandbox built on \texttt{containerd} and \texttt{gVisor}, records the produced artifact hash, and requests a TEE attestation covering the launch measurement, source snapshot hash, and artifact hash. The sandbox matters because the build process itself may execute arbitrary project code: the TEE protects the Enclave Client from the host, and the sandbox protects the Enclave Client from the build it is observing. The authors acknowledge the bootstrap problem, that a verifier needs to know the expected launch measurement of a genuine Enclave Client image, and recommend that the very first such image be produced with reproducible builds, but defer that step to future work.

This line of work changes the verification question. A reproducible build asks whether a second build produces the same bytes. An attestable build asks whether a particular artifact is accompanied by hardware-rooted evidence that a measured build environment observed particular inputs and produced particular output bytes. Both approaches try to close the source-to-binary gap, but they make different tradeoffs. Reproducible builds minimize trust by requiring deterministic reconstruction, while attestable builds reduce reconstruction cost by shifting trust to TEE hardware and its attestation chain.

Kettle~\cite{ref:kettle} builds on this prior work in three specific ways. First, the trust root is the CPU vendor's attestation chain (AMD VCEK, Intel TDX) rather than an operator-controlled PKI, so the signing identity behind every Kettle build is the same one that signs the underlying TEE attestation. Second, Kettle commits an explicit Merkle-rooted input manifest into a SLSA Provenance v1.2 document carried in a canonical in-toto Statement (\S\ref{sec:provenance-format}) and binds the document to hardware via the report-data field, so the evidence format is one that existing supply-chain tooling can already consume. Third, Kettle closes the bootstrap loop that prior work defers: each Kettle release is itself reproducibly built using the Stage\textsuperscript{x} deterministic toolchain~\cite{ref:stagex}, and the CVM image is reproducibly assembled from that Kettle binary and a published recipe (\S\ref{sec:tee-setup}, \S\ref{sec:allow-lists}, \S\ref{sec:kettle-conclusion}). A verifier rebuilds the image from public source, confirms that the resulting launch measurement matches the value published with the release, and adds the measurement to their allow-list. The trust question thereby reduces to source inspection plus the verifier's own toolchain, rather than trust in any image distributor. Kettle also offers an optional pre-attested confidential build flow (\S\ref{sec:confidential-builds}) for sensitive source, which has no equivalent in either system above. The longer-term goal of layering an inner sandbox on top of CVM isolation, in the spirit of~\cite{ref:hugenroth}, is described in \S\ref{sec:open-directions}.

\section{Threat Model}

Every security system has a root of trust, a set of components assumed to be correct because they cannot be verified further down. Attested builds shift where the root of trust sits. In a conventional build, a verifier must trust the entire build infrastructure stack (the cloud provider's software, its operators, other tenants, and the build pipeline itself) to accept that a binary corresponds to its claimed source. Attested builds replace that trust surface with a much smaller one rooted in TEE hardware and a small set of cryptographic primitives.

\subsection{Trust Model}

The trust boundary encompasses the TEE hardware and firmware and a small set of cryptographic primitives. Physical operators are legible attackers. Software-level access by an operator is untrusted, and invasive physical access can break the TEE threat model. The source repository is also trusted in the sense that attestation cannot reason about the intent of the code being built. All other parties are untrusted.

\begin{figure}[H]
\centering
\includegraphics[width=0.92\linewidth]{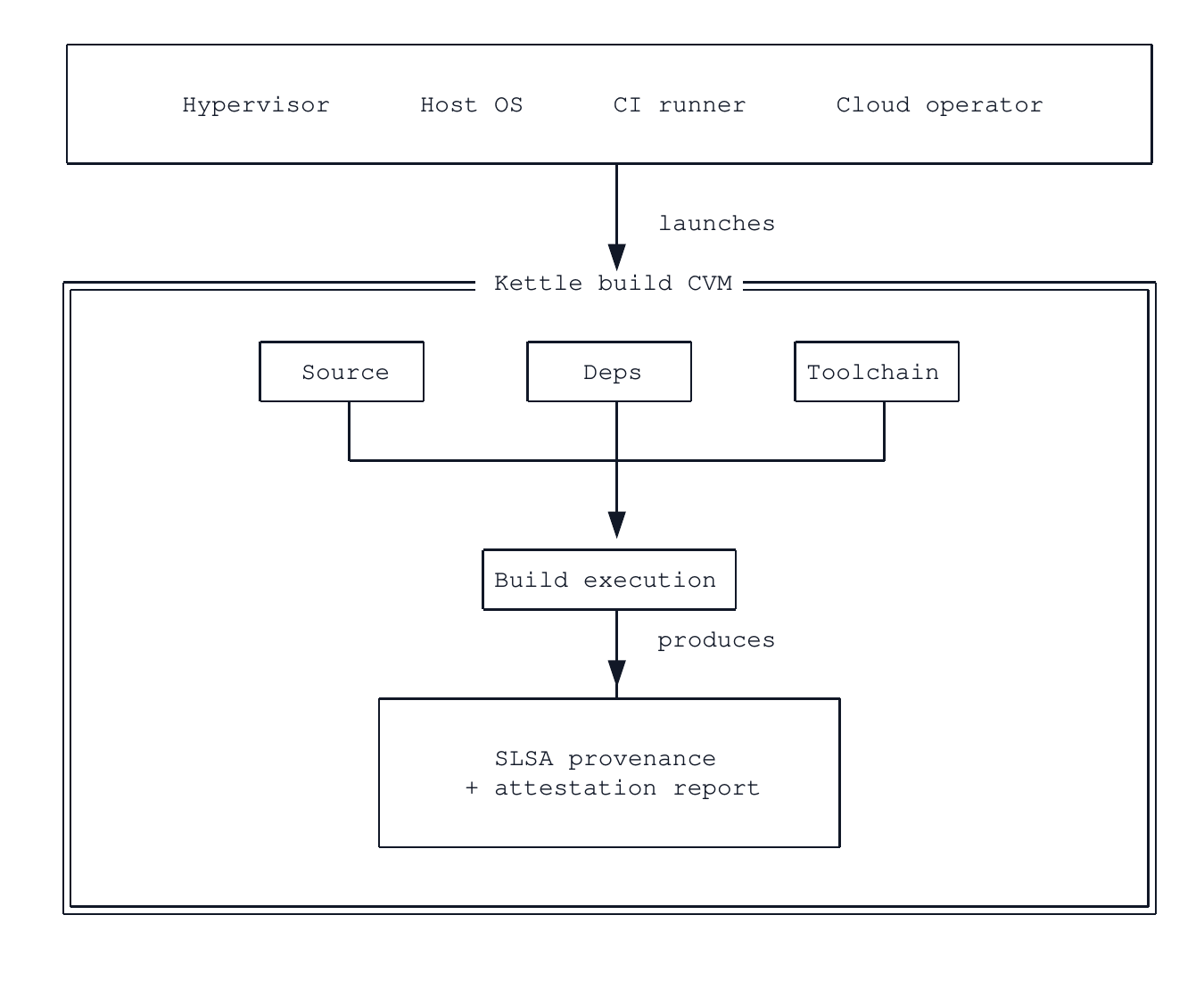}
\caption{The host launches the CVM but cannot observe its memory. Inside the CVM, the build observes its declared inputs and produces a SLSA provenance document together with a hardware-signed attestation report that commits to it. The signed evidence is the only state the verifier later trusts.}
\label{fig:trust-model}
\end{figure}

\begin{table}[H]
\centering
\small
\begin{tabularx}{\linewidth}{@{} X X X @{}}
\toprule
\textbf{Trusted} & \textbf{Untrusted} & \textbf{Out of scope} \\
\midrule
TEE hardware vendor (AMD, Intel) and signing keys & Hypervisor and host OS & Invasive physical attacks by the hardware operator \\
\addlinespace
TEE firmware (security processor microcode) & Cloud provider, datacenter, and CI operators (software access) & CPU probing, memory-bus interposers, chip decapsulation \\
\addlinespace
Cryptographic primitives (AES, ECDSA, SHA) & Other tenants on the same hardware & Physical fault injection against the CPU or memory subsystem \\
\addlinespace
Source repository and package registries (for identity, not intent) & Build orchestration, artifact storage, distribution & \\
\addlinespace
& Network infrastructure & \\
\bottomrule
\end{tabularx}
\caption{Trust model. Hardware vendors and code measured into TEEs are trusted; build orchestration, distribution, and other tenants are not; invasive physical attacks are out of scope.}
\end{table}

Two of these entries are subtle and worth elaborating.

\textbf{Physical operators.} Whoever physically hosts the hardware can attack the TEE using active probing of the memory bus, hardware interposers placed between the CPU and memory, fault injection, or chip decapsulation. Those attacks can break the guarantees of the TEE. They require physical possession of the running machine and specialized equipment, and are outside Kettle's threat model. Software-level access by the same operator (administrators, datacenter technicians, monitoring agents, employees acting through cloud APIs) is untrusted and is in scope. The asymmetry is deliberate. The TEE threat model is scoped against software adversaries with arbitrary host privileges, not against an attacker with invasive physical access.

\textbf{Source and package registries.} The source commit and any registry-fetched dependencies are trusted only for \emph{identity}. The attestation proves that a specific commit and specific packages were used. This prevents package alteration or replacement, and provides a verifiable audit trail, but does not prove that any of them are free of backdoors.

\subsection{Attacks Addressed}

Protection comes from two sources. The first is TEE isolation during the build itself. The second is cryptographic binding between inputs, the build process, and outputs.

\begin{itemize}
\item \textbf{Tampering during the build.} An attacker who compromises the build platform and injects malicious behavior during compilation is normally undetectable. The source is clean but the binary is backdoored. Running the build inside a TEE places the environment outside the host's reach, and the attestation report binds the build to the exact code that was loaded. Any modification to the build process changes the launch measurement and causes verification to fail.
\item \textbf{Tampering after the build.} An attacker with access to artifact storage could substitute a malicious binary for the legitimate one. The provenance produced by an attested build records the digest of every output artifact, and a SHA-256 digest of the provenance document is in turn committed to the TEE attestation report-data field. A substituted artifact fails the digest check against the provenance, and a forged provenance document fails the digest check against the report-data field of the hardware-signed attestation.
\item \textbf{Forged provenance.} Without a hardware root of trust, provenance is a claim. Anyone with a signing key can produce a JSON document asserting that a binary came from a particular source. Attested provenance is bound to a hardware-signed attestation report by committing a SHA-256 digest of the document into the report-data field, with the report itself signed by the TEE platform's attestation key (the AMD VCEK on SEV-SNP, the equivalent on Intel TDX). The certificate chain terminates at the CPU vendor's root of trust, so forging provenance requires producing a valid hardware-signed attestation report whose report-data matches the forged document, which in turn requires compromising the TEE hardware itself.
\item \textbf{Dependency substitution.} A pinned dependency silently swapped at build time fails the checksum recorded in the lockfile and is rejected before the build starts. The provenance records the resolved version and digest of every dependency, so post-hoc substitution is also detectable by any verifier.
\end{itemize}

\subsection{Attacks Not Addressed}

The following are architectural constraints rather than implementation gaps. Stating them explicitly clarifies the actual security posture provided by attested builds.

\begin{itemize}
\item \textbf{Malicious source code.} Attested builds verify that a specific source was used, not that the source is safe. If the upstream repository contains a backdoor, the attested build faithfully reproduces the backdoor and the provenance accurately records the commit that introduced it. Attestation proves identity, not intent.
\item \textbf{Compromised upstream dependencies.} Dependencies fetched from registries are trusted for identity, not for the trustworthiness of their contents. If a registry publishes a malicious package, an attested build that consumes that package produces a verifiable record that the malicious version was used. It does not detect that the package is malicious.
\item \textbf{TEE hardware or firmware compromise.} The trust anchor is the CPU vendor's hardware and key management. If those are compromised, the guarantees break. Firmware vulnerabilities have been found in TEE implementations before and will be found again. The trust surface is substantially smaller than that of a conventional build pipeline, but not zero.
\item \textbf{Invasive physical attacks.} A physical operator with possession of the machine can attack the CPU package, memory bus, or memory subsystem directly. These attacks can break TEE confidentiality or integrity and are outside the software-adversary model Kettle relies on.
\item \textbf{Side-channel attacks.} TEEs share microarchitectural state with the host. Cache timing, branch prediction, and other side channels can leak information. Vendor mitigations have raised the bar significantly, but side channels remain an active research area. Attested builds reduce, but do not eliminate, information leakage during compilation.
\item \textbf{Availability.} The host controls whether the build runs at all. It can refuse to schedule the TEE, terminate it mid-build, or disrupt network connectivity. Attested builds protect confidentiality and integrity, not availability.
\item \textbf{Bugs in the built software.} The TEE protects the execution environment, not the code running in it. If the built application contains a vulnerability, an attacker can exploit it. Attestation proves which code was loaded, not that the code is correct.
\end{itemize}

\section{Architecture}
\label{sec:architecture}

A Kettle build moves a project from ``source on a developer's machine'' to ``artifact accompanied by hardware-rooted evidence'' without trusting anything between those two endpoints. The system has three logical pieces. The \textbf{client} provides a random nonce and a version control reference (such as a git repository and commit ID). The \textbf{host} schedules the build but is otherwise untrusted. The \textbf{Kettle CVM} is a confidential virtual machine that boots from a reproducible Kettle image, runs the build inside the TEE, and emits an evidence bundle verifiable against the CPU vendor's root of trust.

\begin{figure}[H]
\centering
\includegraphics[width=0.98\linewidth]{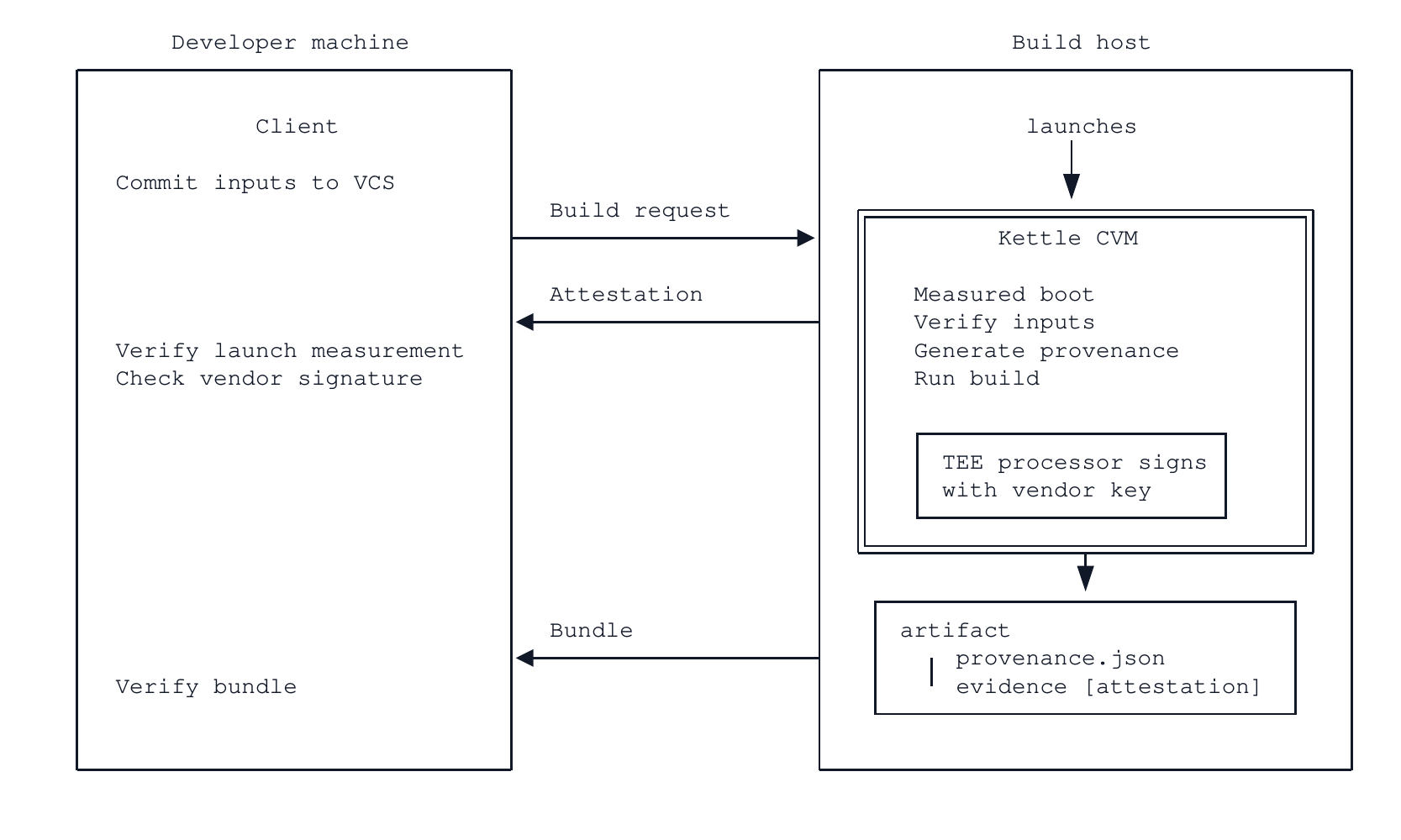}
\caption{Kettle build at a glance. The client commits the input and requests a build. The build runs inside a freshly-launched CVM, which generates a provenance document and emits an attestation signed by the TEE processor with a key chained to the CPU vendor root. The bundle returned to the client is verifiable locally against that root.}
\label{fig:architecture-overview}
\end{figure}

The result is a single bundle (the artifact, \texttt{provenance.json}, and the raw attestation \texttt{evidence}) that any verifier can check locally. The remainder of this section describes each part of that flow in detail.

\subsection{Build Flow}
\label{sec:build-flow}

An attested build runs across four actors. These are the developer's CLI, the host (or CI runner) that schedules the work, the Kettle CVM that performs the build, and the TEE hardware that anchors the attestation. The host can start, stop, and route IO for the CVM, but cannot observe its memory. Everything between the CVM boot and the emitted evidence is shielded by the hardware boundary.

\begin{figure}[H]
\centering
\includegraphics[width=0.98\linewidth]{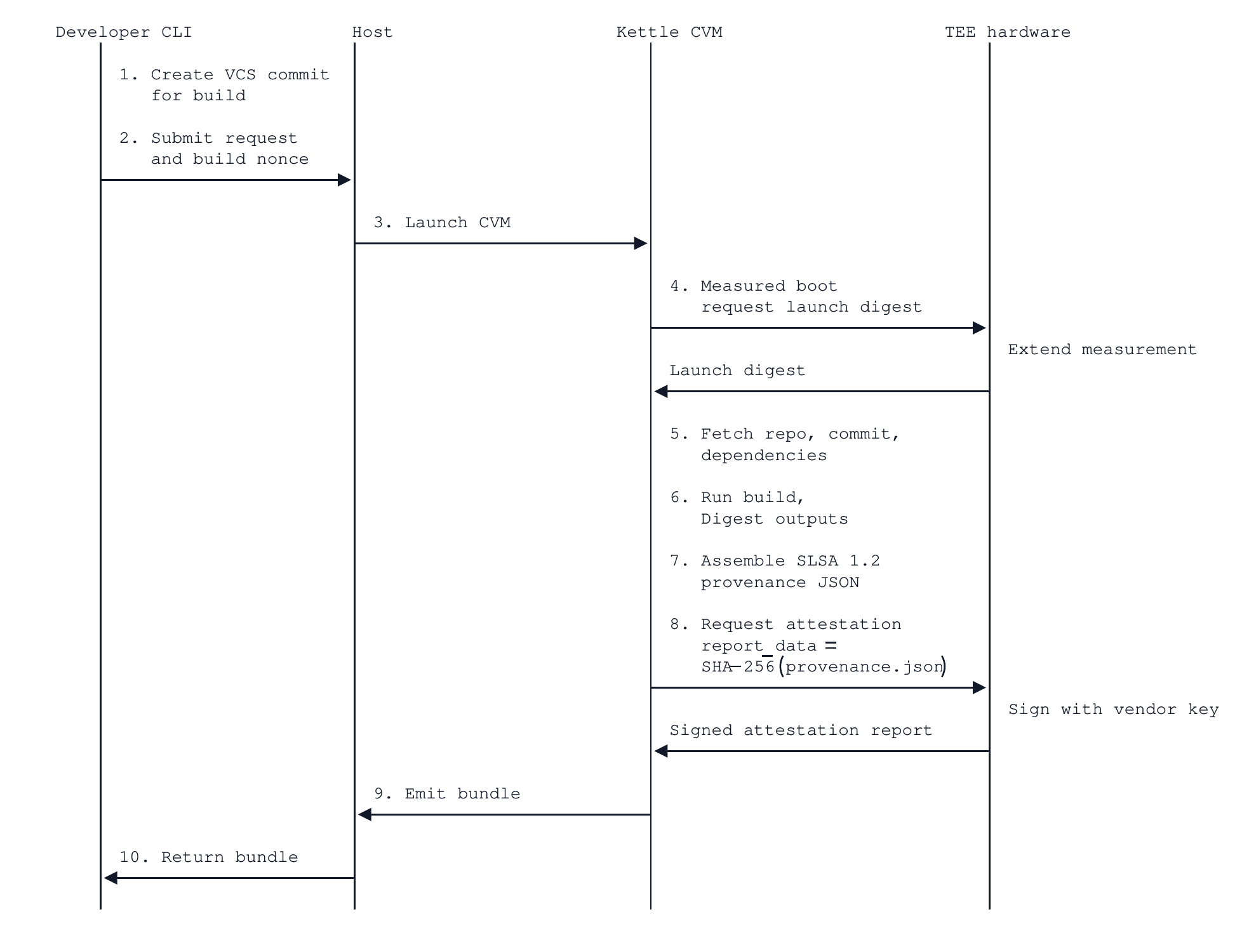}
\caption{Kettle build sequence. The developer creates a VCS commit and submits the build request to the host (1--2). The host launches the CVM (3). Inside the CVM, measured boot extends the launch digest in TEE hardware (4). Kettle fetches the repo, commit, and dependencies, then (5) runs the build and digests outputs (6), and assembles the provenance document as SLSA 1.2 build provenance JSON (7). It then requests an attestation report whose \texttt{report\_data} field contains \texttt{SHA-256(provenance.json)} and the build request nonce (8). The hardware signs the report with the platform attestation key. The artifact and evidence are emitted back to the host (9) and returned to the developer (10).}
\label{fig:build-sequence}
\end{figure}

\S\ref{sec:input-merkle} describes the input manifest committed to the provenance document. \S\ref{sec:tee-setup} covers measured boot and the attestation report. \S\ref{sec:build-execution} covers provenance generation, build execution, output digesting, and how the provenance is committed to the attestation. \S\ref{sec:evidence-chain} specifies the verification procedure. \S\ref{sec:confidential-builds} describes an optional confidential variant of the flow above, in which the requester pre-attests the CVM and sends the source over a TLS channel terminated inside it. The verifier-facing chain is unchanged.

\subsection{Input Merkle Tree}
\label{sec:input-merkle}

Before any build starts, the project's dependencies and source state are locked. This step runs locally, outside the TEE, and produces the input manifest that the CVM will consume.

\textbf{Locking dependencies.} The project must carry a lockfile that pins every dependency to a specific version and a cryptographic digest. This is standard practice in modern package managers, and the lockfile captures the exact dependency graph at a point in time. Ecosystems without first-class lockfiles require the developer to supply an equivalent pinned manifest of resolved versions and digests as part of the build configuration.

\textbf{Input enumeration.} Kettle walks every build input and computes its cryptographic digest:

\begin{itemize}
\item Source code: the git commit identifier, tree digest (a content-addressed digest of the file tree), and repository signature where present.
\item Dependencies: each package identified by name, version, and digest taken from the lockfile.
\item Toolchain: digests of compiler and build-tool binaries.
\end{itemize}

Cached artifacts are checked against their expected digests at this stage, and any mismatch fails the build before the CVM is launched.

\textbf{Merkle tree construction.} All input digests become leaves in a Merkle tree. The tree is constructed in a deterministic order, with git information first, then the lockfile digest, then dependencies in lexicographic order by name, and toolchain digests last. The ordering is fixed by convention so that any third party can reconstruct the same tree from the same inputs. Kettle uses SHA-256 for both leaf and node digests. The Merkle structure (fixed-size inputs concatenated with explicit length prefixes at each internal node) does not admit length-extension attacks against the underlying digest.

\begin{figure}[H]
\centering
\includegraphics[width=0.7\linewidth]{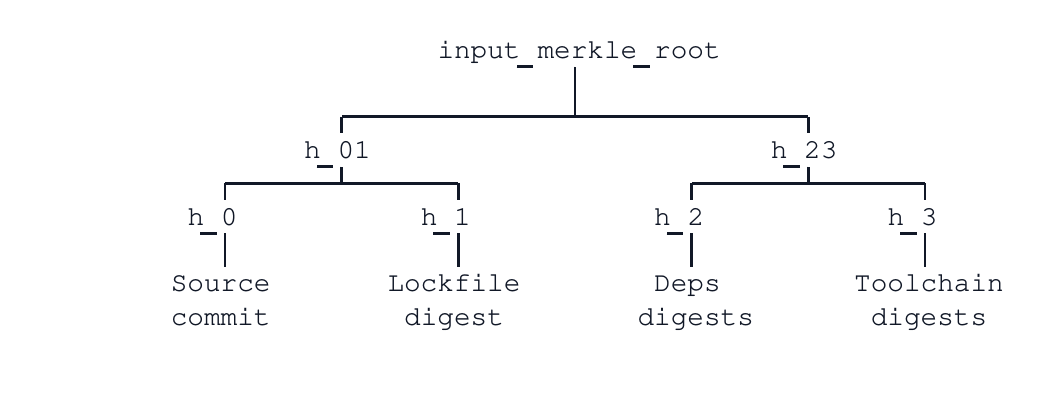}
\caption{Input manifest as a Merkle tree. Each declared input is digested into a leaf, leaves are paired and hashed into internal nodes, and the procedure repeats until a single root remains. The root is recorded in the provenance document, whose digest is committed to the attestation report. Any change to any input bubbles up and changes the root.}
\label{fig:input-merkle}
\end{figure}

The root of this tree is a single digest that uniquely identifies the complete set of build inputs. Any change to any input, even of a single byte, changes the root.

The Merkle structure also enables selective disclosure. To prove inclusion of a specific dependency without revealing the rest of the input set, the builder publishes the path from that dependency's leaf to the root. The verifier checks the path against the published root and confirms that the dependency was part of the build. Selective disclosure matters most when the dependency set is itself sensitive, for example a vendor that does not wish to enumerate proprietary internal packages to an external auditor.

\subsection{TEE Environment Setup}
\label{sec:tee-setup}

The build environment boots inside a TEE with measured boot. From this point onward, every step of the build is bound to evidence rooted in hardware that no software on the host can forge. Concretely, Kettle currently targets confidential VMs implemented by AMD SEV-SNP~\cite{ref:amd-snp} and Intel TDX~\cite{ref:intel-tdx}. The architecture below is described in vendor-agnostic terms but maps directly onto the launch-measurement and attestation primitives those platforms expose.

\textbf{Measured boot chain.} As the TEE initializes, the hardware measures every component that loads. The security processor computes a digest of each component and extends it into a cumulative measurement register, where the extension operation is $\mathrm{new\_measurement} = \mathrm{digest}(\mathrm{old\_measurement} \,\|\, \mathrm{component\_digest})$. Extension is one-way. A measurement register cannot be ``un-extended'', so the final value depends on exactly what was loaded and in what order. The measurement chain covers:

\begin{table}[H]
\centering
\small
\begin{tabularx}{\linewidth}{@{} >{\bfseries}p{0.18\linewidth} X @{}}
\toprule
Component & What gets measured \\
\midrule
Firmware & UEFI/boot code that initializes the TEE \\
Kernel & Kernel image and the kernel command line \\
Initrd & Initial RAM filesystem used to bring up the build environment \\
VM image & The hardened operating system image \\
Kettle & The attested build orchestrator \\
\bottomrule
\end{tabularx}
\caption{Measurement chain components. The cumulative launch measurement extends each loaded layer in order, so the final digest depends on every byte that was loaded and the order it was loaded in.}
\end{table}

Including the kernel command line in the measurement is important. A parameter passed at boot can change kernel behavior (security features, mount options, init binary) without altering the kernel image itself. A measurement chain that omitted the command line would let the host silently change those parameters between builds. Kettle itself is part of this initial measurement. The resulting attestation report identifies the environment as a TEE running a specific Kettle release at a known cumulative measurement, so a verifier can confirm that the entity producing the report is a genuine instance of Kettle rather than arbitrary code making the same claim.

\textbf{Isolation as the primary defense.} By requiring a CVM, Kettle inherits hardware-enforced isolation rather than relying on container-level boundaries. Root access inside the CVM grants no visibility into other build jobs, the host, or other tenants on the same physical hardware, and each build runs in its own fresh CVM, so cross-build contamination is structurally precluded. The base VM image contains only the components required to run Kettle and execute builds, and is built deterministically so that its measurement is predictable across releases.

\textbf{Reproducible CVM image.} Because the CVM image itself does not depend on the inputs of any individual build, it can be produced once per Kettle release as a fully reproducible artifact. Its digest, and the cumulative launch measurement that the platform will report when a CVM boots from it, are public, stable, and tied to a specific Kettle version. Anyone can rebuild the image from the published source and confirm that the resulting launch measurement matches the value pinned in their allow-list (\S\ref{sec:allow-lists}). This is the foundation that lets a verifier recognize a ``genuine Kettle build CVM'' without trusting the operator that hosts it, and it is what enables the optional confidential-input flow described in \S\ref{sec:confidential-builds}.

\textbf{Defence in depth (roadmap).} Several additional hardening layers complement CVM isolation but are not yet implemented in the current Kettle release. These include a mandatory access control policy applied to the build process, system-call filtering that allow-lists only the syscalls needed for compilation and file I/O, and severing network access at the kernel level once inputs are loaded. They are discussed alongside other planned work in \S\ref{sec:open-directions}.

\textbf{Loading inputs.} The CVM consumes inputs in one of two ways. In the simpler mode, Kettle fetches each declared dependency over the network from inside the CVM and only then severs network access. In the stronger mode, all inputs are bundled with the source archive and become part of the initial CVM image, so they are themselves covered by the launch measurement. This eliminates network trust during the build at the cost of a larger image. In both cases, no input is used unless it is part of the commit from the build request, and any mismatch aborts the build before any project code runs.

\textbf{Provenance binding via attestation.} Kettle binds the provenance document to hardware by committing its digest into the attestation report's report-data field. The attestation report itself, signed by the platform attestation key, is the signature on that document. The platform attestation key is provisioned by the hardware vendor and held by the CPU's security processor, so the signing identity behind every Kettle build chains directly to the vendor root of trust. Forging Kettle provenance requires forging an attestation report from genuine TEE hardware.

\textbf{Attestation report.} Once the build has run and the provenance document has been assembled (\S\ref{sec:build-execution}), Kettle requests an attestation report from the platform. The report contains the cumulative launch measurement, platform information (CPU model, firmware version, enabled security features), and a \emph{report-data} field that the guest fills in. Kettle places a SHA-256 digest of the provenance document and the one-time build nonce into the \emph{report-data} field,
\[
  \mathrm{report\_data} = \mathrm{SHA256}(\mathrm{provenance\_document}),\ \mathrm{nonce}
\]
where $\mathrm{provenance\_document}$ is the canonical-JSON serialization of the in-toto Statement described in \S\ref{sec:in-toto}, and the digest covers everything in that document including the input Merkle root, the resolved-dependency list, the output artifact digests, the build metadata, and the build-request nonce supplied by the requester. SEV-SNP and TDX expose a 64-byte report-data field. The SHA-256 digest occupies the leading 32 bytes, and the remaining bytes carry the build-request nonce.

The report signature chains to the CPU vendor's root certificate, so a verifier can confirm in one signature check that the report was produced by genuine TEE hardware, that the measured environment matches an expected Kettle build configuration, and that this report corresponds to one specific provenance document. Any change to the provenance document changes its digest and therefore the report-data the platform signed over.

\begin{figure}[H]
\centering
\includegraphics[width=0.95\linewidth]{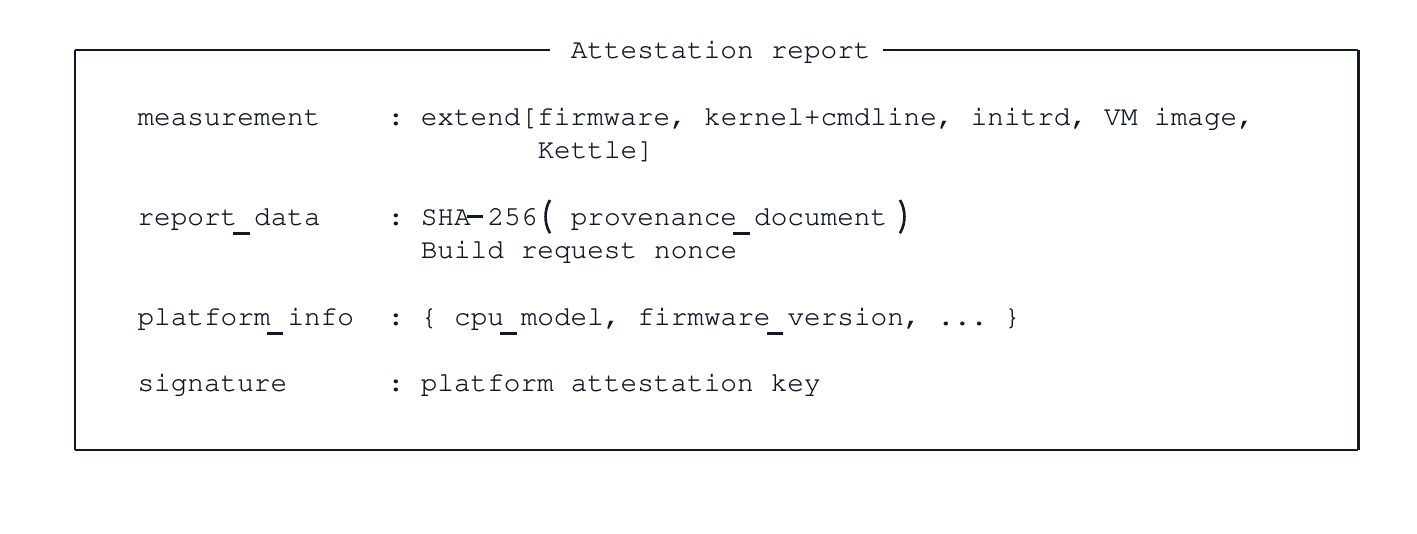}
\caption{Measured boot extends each loaded component into a single cumulative launch measurement, which becomes the \texttt{measurement} field of the attestation report. The \texttt{report\_data} field carries a SHA-256 digest of the canonical provenance document and the nonce from the build request. The whole report is signed by the platform attestation key. The attestation report is therefore the signature on the provenance, with no additional in-guest key required.}
\label{fig:attestation-report}
\end{figure}

The attestation report is the bridge between software claims and hardware proof. The VCS commit determines which inputs will be used, and the attestation proves that a measured Kettle environment used those inputs, ran the build, and produced exactly the provenance document the report-data digest commits to, all within a single hardware-signed object.

\subsection{Build Execution and Provenance}
\label{sec:build-execution}

With the environment measured and inputs verified, the build executes inside the TEE. The output of this stage is a provenance document whose digest is then committed to hardware via the attestation report described in \S\ref{sec:tee-setup}.

\textbf{Build execution.} Kettle invokes the project's existing build toolchain according to the build configuration. Compilation, linking, and packaging proceed exactly as they would on any developer machine, and Kettle does not modify the build process, only the environment it runs in. CVM isolation remains active throughout, and any of the in-CVM hardening layers planned in \S\ref{sec:open-directions} will apply at this stage when introduced.

\textbf{Output digests.} On completion, Kettle computes a SHA-256 digest over each output artifact, and these digests are recorded in the provenance document. Because digest calculation happens inside the TEE, the output digests inherit the same isolation guarantees as the build itself, so a malicious host cannot substitute artifacts between the build finishing and the digests being recorded.

\textbf{Provenance assembly.} Kettle assembles the provenance document inside the CVM. The document is an in-toto Statement carrying an SLSA Provenance v1.2 predicate (\S\ref{sec:provenance-format}) and contains the input manifest (Merkle root and all input digests), the digests of all output artifacts, build metadata, and the build-request nonce supplied by the requester. Kettle then serializes the document under a canonical JSON encoding so that any verifier can recompute the exact byte sequence the platform will sign over. The SHA-256 digest of that canonical serialization is what Kettle places in the report-data field of the attestation request.

\textbf{Hardware-bound provenance.} Provenance integrity comes from the attestation report itself. Once the platform returns a signed report whose report-data matches the SHA-256 digest of the canonical provenance document, the document is bound to the measured Kettle environment. A verifier reconstructs the chain by recomputing the digest from the provenance document, checking it against report-data, and validating the report signature against the CPU vendor's root.

\textbf{Build outputs.} A complete build emits three evidence artifacts alongside the built outputs. \texttt{provenance.json} is the SLSA v1.2 build record carried inside an in-toto Statement, recording inputs, output digests, and build metadata, in the canonical JSON form whose SHA-256 digest the attestation report committed to. \texttt{evidence.json} includes the TEE attestation report whose report-data field carries the provenance digest and build nonce. Together with the source commit, these three artifacts form the complete evidence chain. The provenance binds inputs to outputs, and the attestation binds the whole document to a measured TEE for a specific build request. \S\ref{sec:provenance-format} specifies the format of these artifacts in detail.

\subsection{Evidence Chain and Verification}
\label{sec:evidence-chain}

The result is a cryptographic chain from declared inputs to artifact bytes. Verification walks that chain in order, checking each binding. Each check is meaningful only if the previous check passed.

\begin{figure}[H]
\centering
\includegraphics[width=0.7\linewidth]{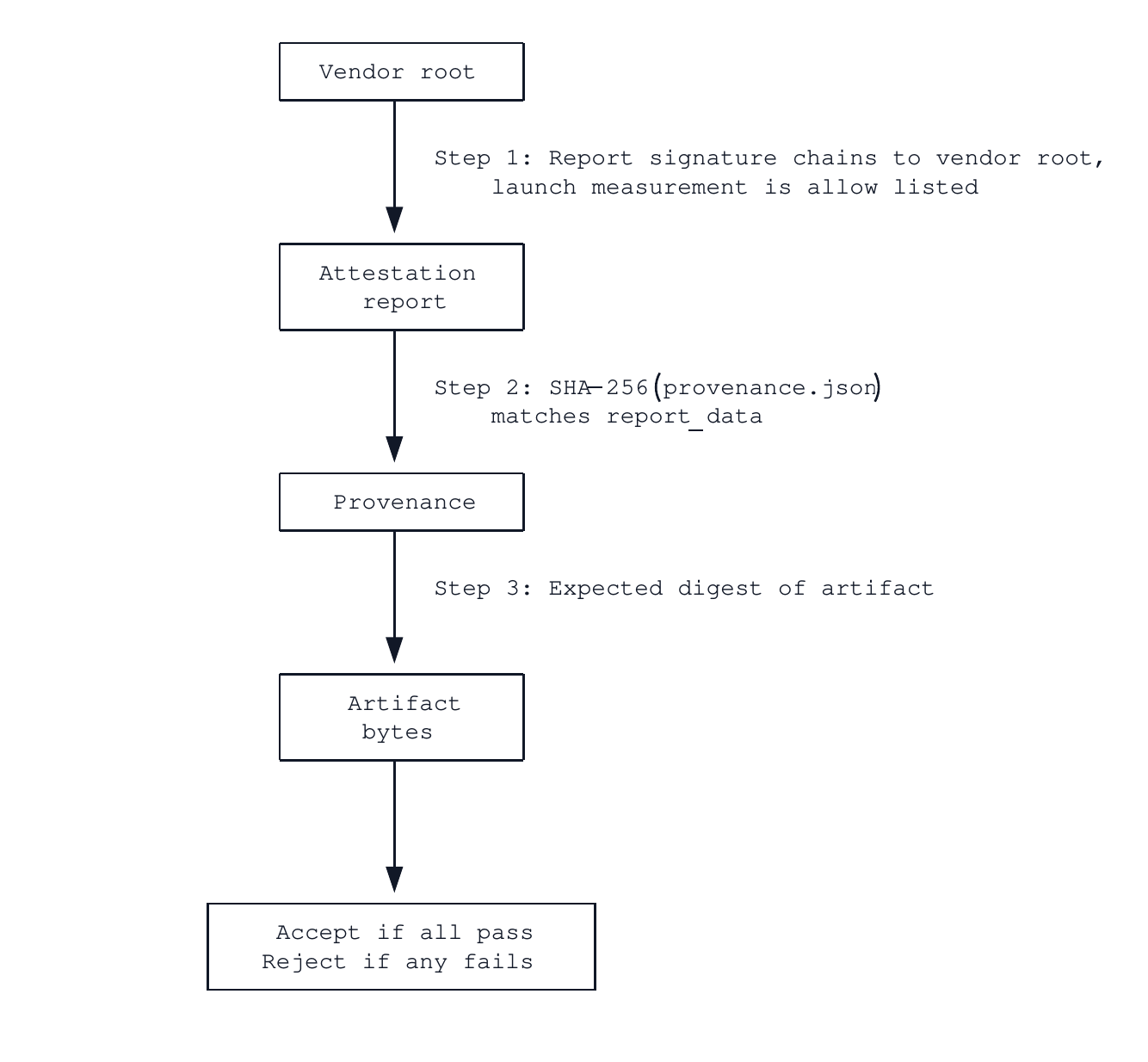}
\caption{Evidence chain and verification. Each link binds to the next, and each numbered check verifies one link. Verification is local. Consumer tooling reads the artifact, \texttt{provenance.json}, and the raw \texttt{evidence} blob from a single bundle and runs the checks in order, with no remote service contacted.}
\label{fig:evidence-chain}
\end{figure}

\begin{table}[H]
\centering
\small
\begin{tabularx}{\linewidth}{@{} >{\bfseries}p{0.30\linewidth} X @{}}
\toprule
Step & What consumer tooling checks \\
\midrule
Step 1: Verify the TEE attestation & The attestation signature chains to the hardware vendor's root of trust, the cumulative launch measurement matches an allow-listed Kettle measurement (\S\ref{sec:allow-lists}), and the freshness nonce embedded in the provenance matches the build-request nonce. \\
\addlinespace
Step 2: Verify the provenance binding & The verifier recomputes $\mathrm{SHA256}(\mathrm{provenance.json})$ over the canonical serialization of the provenance document and checks that it matches the leading 32 bytes of the report-data field. This is the step that turns the attestation report into a signature on the provenance. \\
\addlinespace
Step 3: Verify the artifact binding & The digests of the distributed output artifacts match the digests recorded in the provenance. \\
\bottomrule
\end{tabularx}
\caption{Verification procedure. Each step checks one link in the evidence chain; all must pass for the artifact to be accepted.}
\end{table}

Each link in the chain rejects a corresponding attack:

\begin{table}[H]
\centering
\small
\begin{tabularx}{\linewidth}{@{} >{\bfseries}p{0.32\linewidth} X @{}}
\toprule
Attack & Mechanism that detects it \\
\midrule
Source modified after commit & The tree digest changes, the Merkle root changes, and the provenance file no longer matches the published build. \\
\addlinespace
Dependency substituted before the build & The dependency's digest does not match the lockfile entry, so input verification fails before the build starts. \\
\addlinespace
Build machine compromised & The build runs inside the TEE, so the host cannot read memory or inject code. Loading different inputs changes the provenance commitment or causes input verification to fail. \\
\addlinespace
Malicious Kettle binary & The launch measurement differs from the expected Kettle measurement, and the verifier rejects the attestation. \\
\addlinespace
Output artifact swapped after build & The artifact digest no longer matches the digest recorded in the provenance. \\
\addlinespace
Provenance forged & The SHA-256 digest of the forged provenance does not match the report-data field of the hardware-signed attestation report, so the binding check at step 2 fails. \\
\bottomrule
\end{tabularx}
\caption{Attacks rejected by the evidence chain. Each row maps an attack to the check that detects it.}
\end{table}

When all checks pass, consumer tooling has cryptographic proof that the artifact in hand was produced by a specific build of Kettle running inside attested TEE hardware, over a specific list of inputs, and that every link in that chain is independently verifiable. Verification consists entirely of digest comparisons and one hardware-attestation signature check against the CPU vendor's root, all of which are local operations.

\subsubsection{Measurement Allow-Lists}
\label{sec:allow-lists}

Step 1 above presupposes that a verifier can recognize an allow-listed Kettle measurement. In practice the cumulative launch measurement covers firmware, kernel and command line, initrd, VM image, and Kettle itself, so an allow-list entry is not a single value but a set of acceptable measurements, one per signed Kettle release. A verifier maintains (or imports) a list of these measurements together with the Kettle version, target platform (SEV-SNP, TDX), and any required platform-firmware level it implies. Verification policy is then expressible as ``accept attestations whose measurement appears in the allow-list at version $v$ or later, on platform $p$ with firmware $\geq f$.''

The allow-list itself needs trustworthy origin. Because each Kettle release ships as a reproducible CVM image (\S\ref{sec:tee-setup}), its launch measurement is recomputable from the published source. A verifier, or any independent observer, rebuilds the image and confirms that the resulting measurement matches the value published with the release before adding it to their allow-list. Verifiers that do not wish to track Kettle releases directly can delegate to a third party that publishes a signed allow-list.

\subsection{Confidential Builds}
\label{sec:confidential-builds}

The flow described so far yields \emph{integrity} at the hardware level. A verifier can be sure that the bytes claimed as inputs were the bytes the build observed. It does not, on its own, hide those inputs from the host. In the simple flow the host receives the source archive in plaintext from the developer's CLI and hands it to the CVM at launch, so a malicious operator can still inspect or copy it before the CVM boots.

When the source is itself sensitive (proprietary code, customer data, an unreleased product) the requester can take the additional step of refusing to send any input until the CVM has proven it is a genuine Kettle release, and then sending the input over a confidential channel terminated \emph{inside} the CVM. The reproducible CVM image (\S\ref{sec:tee-setup}) is what makes this practical. Because the launch measurement of a Kettle release is public and stable, the requester can decide, before transmitting any source, whether the CVM they are about to talk to is one they trust.

The flow has three stages and reuses the same \texttt{report\_data} mechanism that already binds inputs and outputs.

\begin{figure}[H]
\centering
\includegraphics[width=0.98\linewidth]{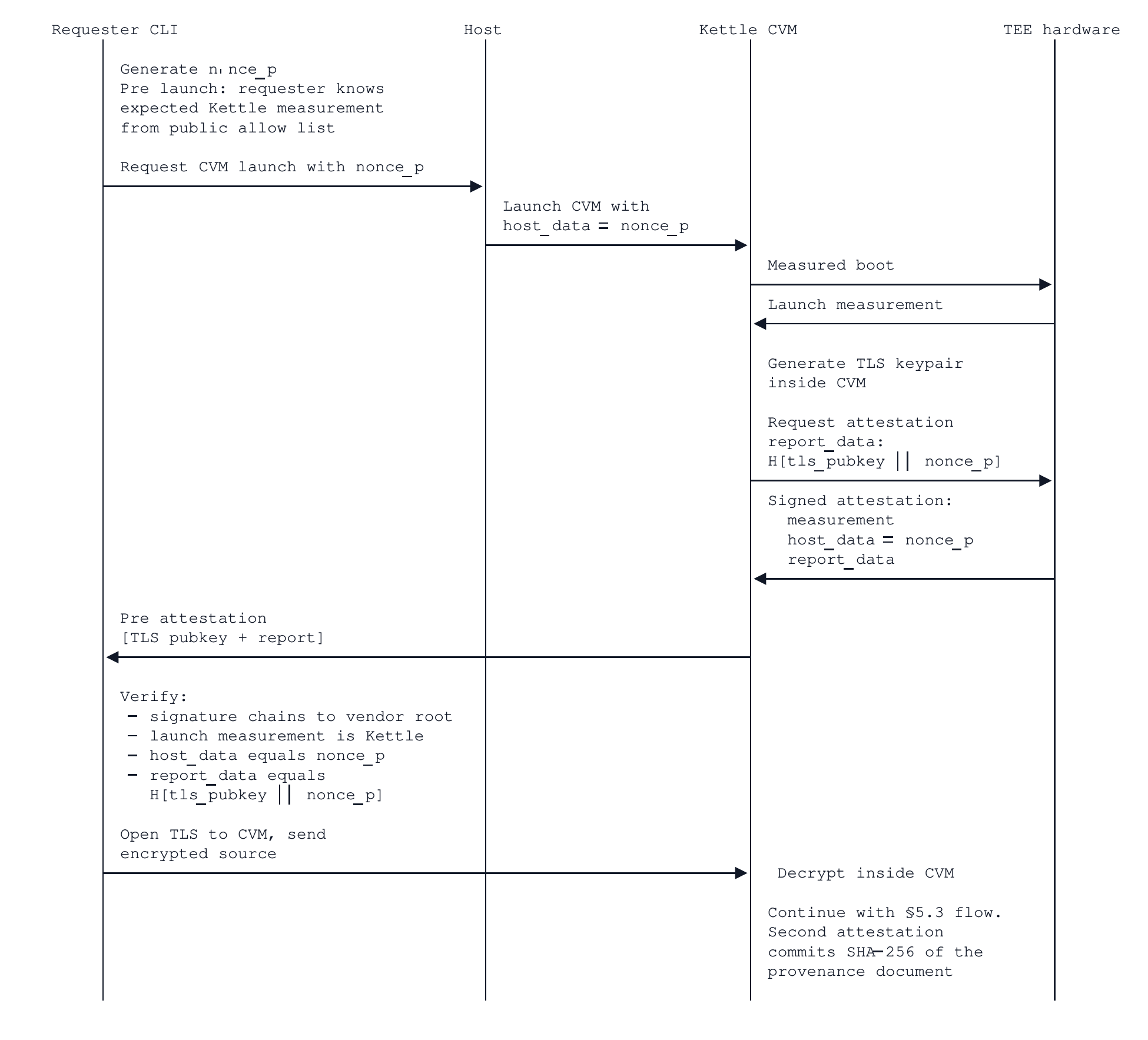}
\caption{Pre-attested confidential build. The requester generates a single nonce that the hypervisor commits into the CVM's \texttt{host\_data} field at launch and that Kettle inside the CVM also includes in \texttt{report\_data}. The first attestation therefore proves three things at once. The CVM is running Kettle code, it was uniquely launched for this requester, and the report itself is fresh. Source is then delivered over a TLS channel terminated inside the CVM, and the standard \S\ref{sec:tee-setup} attestation over the provenance document follows.}
\label{fig:confidential-build}
\end{figure}

\textbf{Stage 1, pre-attestation.} The requester generates a fresh $\mathrm{nonce}_p$ and submits it with the CVM-launch request. The hypervisor writes $\mathrm{nonce}_p$ into the CVM's \texttt{host\_data} field, which the platform records before the guest boots and includes in every attestation report the CVM produces. Once booted, the CVM generates a TLS keypair inside its memory and requests an attestation report whose \texttt{report\_data} field carries $\mathrm{SHA256}(\mathrm{tls\_pubkey} \,\|\, \mathrm{nonce}_p)$. The report is returned to the requester along with the TLS public key. The TLS keypair is used only to terminate the channel inside the CVM and plays no role in signing provenance.

\textbf{Stage 2, verification before transmission.} The requester checks the attestation locally. The signature must chain to the CPU vendor's root of trust, the launch measurement must match an allow-listed Kettle release on the expected platform, \texttt{host\_data} must equal the $\mathrm{nonce}_p$ the requester just submitted, and \texttt{report\_data} must match $\mathrm{SHA256}(\mathrm{tls\_pubkey} \,\|\, \mathrm{nonce}_p)$ for the TLS public key just received. Each check answers a different freshness question. The launch measurement covers code identity, \texttt{host\_data} covers CVM uniqueness (this is a freshly launched VM provisioned for this requester and not a replay or shared instance), and \texttt{report\_data} covers attestation freshness plus the channel binding. Because the Kettle CVM image is reproducible and its measurement is published with each release, the requester can perform every one of these checks without trusting the operator that hosts the CVM. If any check fails, the requester walks away without ever revealing the source.

\textbf{Stage 3, confidential delivery and build.} The requester opens a TLS session to the CVM using the verified public key as the channel endpoint, and transmits the source archive over that session. The host can observe ciphertext and the launch measurement, but cannot read the inputs. Inside the CVM, Kettle decrypts the inputs, and proceeds with the standard build flow (\S\ref{sec:tee-setup}, \S\ref{sec:build-execution}). A \emph{second} attestation report is produced after the build, with $\mathrm{report\_data} = \mathrm{SHA256}(\mathrm{provenance\_document})$ exactly as in the non-confidential flow. The two attestations share a launch measurement (the same Kettle CVM produced both) and chain together. The first proves the channel, the second proves the build.

This flow does not change the verifier-facing evidence chain at all. Verifiers consume the same \texttt{provenance.json} and \texttt{evidence} files and run the same checks (\S\ref{sec:evidence-chain}). What it changes is the \emph{requester-facing} trust posture. Confidentiality of the source is enforced cryptographically rather than by trusting the host, so a build can run inside infrastructure the requester would otherwise be unwilling to send code to.

The confidential flow is opt-in. Public open-source builds can use the simpler path of \S\ref{sec:build-flow} with no loss of integrity guarantees, and only requesters with sensitive inputs incur the extra round trip and the additional pre-launch verification step.

\section{Provenance Format and Standards}
\label{sec:provenance-format}

The evidence chain of \S\ref{sec:architecture} is only useful if its artifacts can be consumed by tools other than Kettle. A proprietary format would put attested builds in a silo. Only Kettle's own verifier could read them, and the rest of the supply-chain ecosystem (registries, scanners, deployment systems, audit tools) would have to integrate against a custom schema. Kettle therefore expresses provenance using two existing standards. The \textbf{in-toto attestation framework}~\cite{ref:in-toto} provides the outer signed-statement structure, and \textbf{SLSA Provenance v1.2}~\cite{ref:slsa-provenance} is the predicate carried inside it. This section describes how those formats are populated and how they map onto the SLSA Build levels.

\subsection{in-toto Statement Layer}
\label{sec:in-toto}

in-toto defines a generic envelope for statements about software artifacts. A statement names a set of \emph{subjects} (the artifacts the claim is about, identified by digest), declares a \emph{predicateType} (the schema of the claim), and carries the claim itself in the \emph{predicate} field. Rather than wrapping the Statement in a separate signature envelope, Kettle commits the SHA-256 digest of its canonical serialization into the attestation report's \texttt{report\_data} field, so the hardware-signed report is itself the binding evidence on the Statement. The concrete shape of a Kettle Statement is shown in \S\ref{sec:slsa-predicate}.

Subjects are identified by content digest, not filename, so a Statement is bound to specific bytes rather than a reusable label. The predicate type is a URI identifying the schema by which the predicate field should be interpreted. The framework is deliberately extensible. SBOMs, code-review attestations, test-result attestations, and vulnerability scans all reuse the same outer structure with different predicate types, and a single artifact can accumulate multiple in-toto Statements over its lifecycle. Kettle emits one such Statement per build, with the SLSA Provenance v1.2 predicate type, and binds it to hardware via the attestation report described in \S\ref{sec:tee-setup}.

\subsection{SLSA Provenance Predicate}
\label{sec:slsa-predicate}

The SLSA Provenance v1.2 predicate has two top-level sections: \texttt{buildDefinition} describes what was built, and \texttt{runDetails} describes how the build executed. An abbreviated example illustrates the fields Kettle populates:

\begin{lstlisting}[language=json]
{
  "_type": "https://in-toto.io/Statement/v1",
  "subject": [
    { "name": "my-app",
      "digest": { "sha256": "1d1ea25c371d4f6de8d6e3c26fdad2238..." } }
  ],
  "predicateType": "https://slsa.dev/provenance/v1",
  "predicate": {
    "buildDefinition": {
      "buildType": "https://kettle.confidential.ai/cargo-build/v1",
      "externalParameters": {
        "repository": "https://github.com/org/repo",
        "ref": "refs/heads/main"
      },
      "internalParameters": {
        "tee_platform": "sev-snp",
        "kettle_version": "0.4.0"
      },
      "resolvedDependencies": [
        { "uri": "git+https://github.com/org/repo@refs/heads/main",
          "digest": { "gitCommit": "a1b2c3d4..." } },
        { "uri": "pkg:cargo/serde@1.0.228",
          "digest": { "sha256": "9a8e94ea..." } }
      ]
    },
    "runDetails": {
      "builder": { "id": "https://kettle.confidential.ai/tee-builder/v1" },
      "metadata": {
        "invocationId": "build-12345",
        "startedOn":  "2026-01-15T10:30:00Z",
        "finishedOn": "2026-01-15T10:35:00Z"
      }
    }
  }
}
\end{lstlisting}

Three fields carry most of the security-relevant content:

\begingroup\sloppy
\begin{itemize}
\item \textbf{\texttt{buildType}.} A URI that fixes the schema of \texttt{externalParameters} and \texttt{internalParameters}. Different toolchains (Cargo, Nix, Bazel) and different build configurations get different build types, so verifiers can apply build-type-specific policy without parsing free-form fields.
\item \textbf{\texttt{externalParameters} vs.\ \texttt{internalParameters}.} External parameters are inputs supplied from outside the trusted control plane (which repository to build, which ref to resolve, which entry point to invoke), and a verifier \emph{must} check these against expectations because they are attacker-influenceable. Internal parameters are set by Kettle itself inside the CVM (target platform, Kettle release, build-time configuration) and are trusted because the platform that emits them is trusted. Pushing build configuration into the source tree (where it is covered by the source commit digest) rather than into \texttt{externalParameters} reduces what a verifier has to police.
\item \textbf{\texttt{resolvedDependencies}.} Every input fetched during the build, recorded as a \texttt{ResourceDescriptor} with URI, digest, and (optionally) name and download location. Dependencies use Package URLs (PURLs), e.g.\ \texttt{pkg:\allowbreak cargo/\allowbreak serde@1.0.228}, so verifiers can cross-reference them against registries or vulnerability databases without bespoke parsers. The distinction between \texttt{externalParameters} (what was \emph{requested}) and \texttt{resolvedDependencies} (what was \emph{actually fetched}) is load-bearing: a request for \texttt{refs/heads/main} resolves to a specific commit, and the resolution is what the attestation witnesses.
\item \textbf{\texttt{builder.id}.} A URI identifying the build platform. Crucially, this represents the \emph{transitive closure} of everything a verifier is trusting to faithfully run the build and emit accurate provenance, including the TEE platform, the Kettle release, and the measurement allow-list policy. Verifier policy is typically expressed as ``accept builds with \texttt{builder.id} $X$ at SLSA Build level $L$.''
\end{itemize}
\endgroup

\subsection{SLSA Build Levels}

SLSA's Build track defines three levels of supply-chain integrity, summarised in~\cite{ref:slsa-provenance}:

\begin{table}[H]
\centering
\small
\begin{tabularx}{\linewidth}{@{} >{\bfseries}p{0.08\linewidth} >{\bfseries}p{0.20\linewidth} X @{}}
\toprule
Level & Summary & Key requirements \\
\midrule
L1 & Provenance exists & Each artifact has provenance describing how it was built. May be unsigned. Catches mistakes, trivial to forge. \\
\addlinespace
L2 & Hosted build platform & Provenance is signed by a hosted build platform. Forging requires an explicit attack rather than a configuration error. \\
\addlinespace
L3 & Hardened builds & Build platform has strong tamper-resistance. Builds are isolated from one another. Signing keys are inaccessible to user-defined build steps. \\
\bottomrule
\end{tabularx}
\caption{SLSA Build track levels.}
\end{table}

Attested builds, as implemented by Kettle, target SLSA Build L3 with hardware enforcement of the L3 requirements:

\begin{table}[H]
\centering
\small
\begin{tabularx}{\linewidth}{@{} >{\bfseries}p{0.32\linewidth} X @{}}
\toprule
L3 requirement & How Kettle satisfies it \\
\midrule
Provenance generated by build platform's trusted control plane & The Kettle orchestrator runs in a separate VM on the host and launches a fresh CVM for each build, so the user-supplied build never executes alongside the control plane and cannot interfere with provenance assembly (\S\ref{sec:build-execution}). \\
\addlinespace
Provenance signed by build platform & The SHA-256 digest of the provenance is committed to the attestation report-data field, and the report is signed by the TEE platform attestation key (e.g.\ AMD VCEK) which chains to the CPU vendor's root of trust. \\
\addlinespace
Builds isolated from one another & Each build runs in its own fresh CVM, and cross-build interference is precluded by hardware memory encryption rather than by soft policy. \\
\addlinespace
Signing keys inaccessible to user-defined build steps & There is no in-guest signing key. The platform attestation key is held by the CPU's security processor and is unreachable from any guest software, including user-supplied build steps. \\
\bottomrule
\end{tabularx}
\caption{How Kettle satisfies SLSA Build L3 with hardware enforcement.}
\end{table}

Hardware enforcement of these requirements is the substantive difference from a typical L3 implementation. The build platform's signing identity is not a software key sitting on a build host but the platform attestation key, held in silicon and reachable only via the platform's attestation interface. A compromised build script, a privileged operator on the host, and a malicious other tenant on the same machine all face the same wall.

\subsection{Build Outputs Recap}

Each Kettle build emits three artifacts whose roles map onto the format above:

\begingroup\sloppy
\begin{itemize}
\item \textbf{\texttt{provenance.json}}. The in-toto Statement described above, with the SLSA Provenance v1.2 predicate, serialized in canonical JSON. Contains the source commit and tree digest, the lockfile digest, the toolchain digests, and the \texttt{input\_merkle\_root} over all of them. This file's SHA-256 digest is what the attestation report's report-data field commits to.
\item \textbf{\texttt{evidence}}. The raw TEE attestation report. The report-data field carries \texttt{SHA-256(provenance.json)} (over the canonical JSON form), and the signature chains to the CPU vendor's root of trust. This is what roots the rest of the chain in hardware.
\end{itemize}
\endgroup

\section{Conclusion}

Attested builds make it possible to definitively connect a specific executable with the inputs used to create that executable without requiring bit-for-bit reproducibility. By binding exact build inputs, a measured build environment, the assembled provenance document (via its digest in the attestation report-data field), and exact build outputs into one evidence chain, a verifier can evaluate a binary without trusting the build host or re-running the build.

\subsection{Trust Surface Comparison}

The shift from a conventional build pipeline to attested builds is best understood as a change in what a verifier must trust to accept that a binary corresponds to its source. In a conventional pipeline, that trust extends across the entire build infrastructure, its operators, and its distribution channel. Verification is policy-based, and a consumer of the binary trusts that procedures were followed, that access controls were configured correctly, and that credentials were not compromised at any point along the chain.

In the attested model, the trust surface contracts to the TEE hardware vendor, the TEE firmware, and the source infrastructure (for input identity). The build infrastructure and distribution channel move outside the trust boundary. Physical operators are not trusted for invasive attacks. They are treated as software adversaries when they act through host privileges, cloud APIs, monitoring agents, or operator access, and as out-of-scope attackers when they physically probe or modify the machine. Verification is cryptographic. The hardware attestation is checked against the vendor root, the provenance document is checked against the digest in the attestation report-data field, and artifact digests are checked against the provenance. The smaller, well-defined trust surface replaces a broad, opaque one.

\begin{table}[H]
\centering
\small
\begin{tabularx}{\linewidth}{@{} >{\bfseries}p{0.27\linewidth} X X @{}}
\toprule
Component & Conventional build & Attested build \\
\midrule
CI runner & trusted & untrusted \\
Build host & trusted & untrusted \\
Release signing key & trusted & replaced by TEE attestation key (vendor-rooted) \\
Registry / distribution & trusted & untrusted \\
Physical operator & trusted & software access untrusted, invasive attacks out of scope \\
TEE hardware & not in trust path & trusted \\
Hardware vendor root & not in trust path & trusted \\
Source identity & trusted & trusted \\
\bottomrule
\end{tabularx}
\caption{What a verifier must trust. The attested model contracts the trust surface to TEE hardware and the vendor root, while the build infrastructure and distribution channel move outside the boundary.}
\end{table}

\subsection{Open Directions}
\label{sec:open-directions}

\begin{itemize}
\item \textbf{In-CVM defense in depth.} Current Kettle releases rely on CVM isolation and a minimal VM image as the primary boundary around the build process. Several additional layers, discussed in \S\ref{sec:tee-setup}, are planned but not yet implemented: a mandatory access control policy applied to the build process so that root inside the CVM does not imply unrestricted file access, system-call filtering that allow-lists only the syscalls needed for compilation and file I/O, and severing network access at the kernel level once inputs are loaded. None of these are required for the cryptographic chain to hold, but each shrinks the attack surface available to a malicious build script or compromised toolchain \emph{inside} the CVM.
\item \textbf{Inner sandbox around the build process.} A complementary direction, in the spirit of Hugenroth et al.~\cite{ref:hugenroth}, is to run the build in an inner sandbox even within the CVM, separating the trusted Kettle observer (which records inputs and produces provenance) from the untrusted project code (which compiles and links). The CVM protects Kettle from the host, and the sandbox would protect Kettle from the build it is observing.
\item \textbf{Digest algorithms.} The Merkle tree of inputs and the report-data commitment currently use SHA-256 with explicit length prefixes at each internal node, which precludes length-extension attacks against the underlying digest. There are no known weaknesses in this construction, but SHA-3 or Blake3 would remove length-extension as an algorithmic concern at all and provide better performance on long inputs.
\item \textbf{Per-file input digests.} Git digests files, trees, and commits using SHA-1, which has documented practical collision attacks (the SHAttered work and successors). Kettle reduces the practical impact by recording both the commit digest and the tree digest in the provenance, so an attacker would need to manufacture a SHA-1 collision in two coupled values simultaneously. Enumerating and digesting each file in the source repository under SHA-2, or better yet SHA-3, would close this residual exposure entirely, at the cost of additional provenance construction time on large source trees.
\end{itemize}

\subsection{Kettle}
\label{sec:kettle-conclusion}

\href{https://github.com/lunal-dev/kettle}{Kettle} is an open-source implementation of attested builds~\cite{ref:kettle}. By committing the SHA-256 digest of each build's provenance document into the TEE attestation report and recording every input, output, and environment measurement in an in-toto Statement carrying an SLSA Provenance v1.2 predicate, Kettle provides supply-chain assertions of comparable strength to bit-for-bit reproducible builds without requiring deterministic compilation.

Kettle releases are themselves fully reproducible, built on top of the Stage\textsuperscript{x} deterministic toolchain~\cite{ref:stagex} and published into the project repository. Any release of Kettle can be independently verified by cloning the repo and executing the reproducible build script, creating a new binary with the same checksum.

% =============================================================================
% References
% =============================================================================


\begin{thebibliography}{99}

\bibitem{ref:reproducible-builds}
Reproducible Builds.
\newblock \emph{Reproducible Builds project}.
\newblock \url{https://reproducible-builds.org}.

\bibitem{ref:lamb-zacchiroli}
C.~Lamb and S.~Zacchiroli.
\newblock Reproducible Builds: Increasing the Integrity of Software Supply Chains.
\newblock \emph{IEEE Software}, 39(2):62--70, 2022.
\newblock \url{https://doi.org/10.1109/MS.2021.3073045}. Preprint: \url{https://arxiv.org/abs/2104.06020}.

\bibitem{ref:nix}
E.~Dolstra, M.~de~Jonge, and E.~Visser.
\newblock Nix: A Safe and Policy-Free System for Software Deployment.
\newblock In \emph{Proceedings of the 18th USENIX Conference on System Administration (LISA '04)}, pp.~79--92, 2004.

\bibitem{ref:tee-compile}
Automata Network.
\newblock \emph{Creating Attestable Builds with AWS Nitro Enclaves}.
\newblock AWS Builder Center, September 2024.
\newblock \url{https://builder.aws.com/content/2lZdD8iDfAGbQpWK8g8fGvQ1kp8/creating-attestable-builds-with-aws-nitro-enclaves}.

\bibitem{ref:hugenroth}
D.~Hugenroth, M.~Lins, R.~Mayrhofer, and A.~R. Beresford.
\newblock Attestable Builds: Compiling Verifiable Binaries on Untrusted Systems using Trusted Execution Environments.
\newblock In \emph{Proceedings of the 2025 ACM SIGSAC Conference on Computer and Communications Security (CCS '25)}, pp.~4514--4528, ACM, 2025.
\newblock \url{https://doi.org/10.1145/3719027.3765128}. Preprint: \url{https://arxiv.org/abs/2505.02521}.

\bibitem{ref:amd-snp}
AMD.
\newblock \emph{AMD SEV-SNP: Strengthening VM Isolation with Integrity Protection and More}.
\newblock Document 70366, 2020.
\newblock \url{https://docs.amd.com/v/u/en-US/SEV-SNP-strengthening-vm-isolation-with-integrity-protection-and-more}.

\bibitem{ref:intel-tdx}
Intel.
\newblock \emph{Intel Trust Domain Extensions (Intel TDX) Module Base Architecture Specification}.
\newblock Document ID 853294, 2025.
\newblock \url{https://www.intel.com/content/www/us/en/content-details/853294/intel-trust-domain-extensions-intel-tdx-module-base-architecture-specification.html}.

\bibitem{ref:in-toto}
S.~Torres-Arias, H.~Afzali, T.~K. Kuppusamy, R.~Curtmola, and J.~Cappos.
\newblock in-toto: Providing farm-to-table guarantees for bytes and bits.
\newblock In \emph{Proceedings of the 28th USENIX Security Symposium}, pp.~1393--1410, 2019.
\newblock \url{https://www.usenix.org/conference/usenixsecurity19/presentation/torres-arias}. Specification at \url{https://github.com/in-toto/attestation}.

\bibitem{ref:stagex}
Stage\textsuperscript{x}.
\newblock \emph{A container-native, full-source bootstrapped, reproducible toolchain}.
\newblock \url{https://stagex.tools}, source at \url{https://codeberg.org/stagex/stagex}.

\bibitem{ref:slsa-provenance}
SLSA Community.
\newblock \emph{Build: Provenance v1.2}.
\newblock \url{https://slsa.dev/spec/v1.2/build-provenance}, 2025.

\bibitem{ref:kettle}
Lunal Dev.
\newblock \emph{Kettle: attested builds for verifiable software provenance}.
\newblock \url{https://github.com/lunal-dev/kettle}.

\end{thebibliography}
\end{document}